\documentclass[aps,prl,reprint,groupedaddress]{revtex4-2}

\usepackage{graphicx}
\usepackage{dcolumn}
\usepackage{bm}
\usepackage[inkscapearea=page]{svg}

\usepackage{xcolor}
\usepackage{xr}
\makeatletter

\newcommand*{\addFileDependency}[1]{
\typeout{(#1)}
\@addtofilelist{#1}
\IfFileExists{#1}{}{\typeout{No file #1.}}
}\makeatother

\newcommand*{\myexternaldocument}[1]{%
\externaldocument{#1}%
\addFileDependency{#1.tex}%
\addFileDependency{#1.aux}%
}

\myexternaldocument{supplemental_material}

\begin{document}

\preprint{APS/123-QED}

\title{Flow-driven Stretch Fluctuations Cause Anomalous Rate-Thinning In Elongating Associative Polymers}

\author{Songyue Liu, Thomas C. O'Connor}
\affiliation{%
 Department of Materials Science and Engineering, Carnegie Mellon University, Pittsburgh, PA, USA
}%

\date{\today}

\begin{abstract}

We use nonequilibrium molecular dynamics simulations to verify recent tube-model predictions that associative polymer networks exhibit broad stretch fluctuations during elongational flow. Simulations further show that these fluctuating dynamics give rise to the rate-dependent extensional viscosity $\eta_E$ measured in filament stretching experiments on H-bonding networks. Simulations model bivalent associative networks with a reactive bead-spring model for varying association strength and extensional strain rate. We observe that stretch fluctuations are driven by a new form of chain tumbling, where chains continually collapse and elongate as their associations break and reform within the convecting network. This produces a broad, nearly uniform distribution of chain stretch over a wide range of strain rates, manifesting as a rate-independent plateau in the extensional stress. Our results show that the nonlinear viscoelasticity of associative networks is dominated by large fluctuations in molecular response, which cannot be captured by current mean-field models. 

\end{abstract}

\maketitle

Associative polymer networks (APNs) are self-assembled networks that form when polymer chains interconnect through a variety of transient and reconfigurable chemical bonds \cite{Wu2021,Zhang2018DynamicsPolymers}. 
Unlike conventional thermosets, APN networks can constantly rearrange their network topology and can adapt in response to stimuli like flow or environmental conditions \cite{Binder2006SupramolecularSide-Chain, Wemyss2021DynamicMachines}.
Bugs and insects routinely exploit this adaptivity to produce associative silk fibers and webbing, and many researchers wish to replicate these biological materials to create novel and sustainable polymers \cite{Wemyss2021DynamicMachines, Webber2022DynamicInteractions, Guo2020EngineeringProcessing,Volkov2015OnReview}.
Extraordinarily, the linear viscoelasticity of APNs shows broad chemical universality, such that the linear response of many different associative chemistries can be captured by the same renormalized tube models \cite{Baxandall1989DynamicsChains,Leibler1991DynamicsNetworks,Rubinstein2001DynamicsPolymers}.
However, this simple picture does not extend to highly nonlinear flows where APNs display novel physics and flow behavior that are difficult to predict and control \cite{MahmadRasid2020UnderstandingShear}.

Filament stretching experiments of various APN chemistries have shown that they display substantial strain hardening relative to unassociative melts \cite{Shabbir2015EffectMelts, Lopez-Barron2024MicrostructureVitrimers, Lopez-Barron2022ExtensionalIonomers, Lopez-Barron2024MicrostructureVitrimers,Wu2018MolecularIonomers,Shabbir2017NonlinearMelts,Ling2012LinearMelts}.
This enhanced strain hardening implies that associative bonds slow chain relaxation and thus enhance chain stretching, which is consistent with mean-field ideas for linear-response.
However, experiments have also observed APNs to display a dramatic rate-thinning of the extensional viscosity $\eta_E$ \cite{Shabbir2015EffectMelts}
with systems of varying association density all displaying near-perfect rate-thinning $\eta_E\sim\dot{\epsilon}^{-1}$, indictative of an emergent rate-independent plateau stress $\sigma_E^*$.
Some new physics drives the emergence of this stress scale, which is not anticipated by current models or observed in unassociative melts that usually display a thinning exponent of $-1/2$ \cite{Huang2022WhenRheology,Costanzo2016ShearEntanglements}.

In a recent Letter, Schaefer and McLeish proposed that the physics governing APN viscoelasticity could be large fluctuations in chain stretch during flow \cite{Schaefer2021PowerFlow}.
Their theory predicts that flow drives convective rearrangement of associative bonds, causing chains to develop broad distributions of chain stretch such that a small fraction of highly stretched chains can dominate the nonlinear rheology - in violation of mean-field approximations.
In later studies, Schaefer and McLeish argue that it is precisely these stretch fluctuations that enable silkworms to spin highly crystalline silk fibers in ambient conditions \cite{Schaefer2021StretchingFlow, Schaefer2022TheoreticalCrystallization}.
If true, then understanding the physics of associative stretch fluctuations could be the key to discovering better bioinspired methods for processing sustainable polymers.

Here we apply nonequilibrium molecular dynamics simulations of coarse-grained APNs \cite{Liu2024AEquilibrium} to characterize the coupling between bond dynamics and nonlinear extensional flows.
Our simulations reproduce the macroscopic strain hardening and rate-thinning of $\eta_E$ observed in experiments on H-bonding polyacrylates \cite{Shabbir2015EffectMelts}, and directly relate these trends to large fluctuations in chain stretch that are consistent with the theoretical predictions of Schaefer and McLeish.
We observe that both features are mediated by a new form of nonequilibrium chain dynamics in which associative chains tumble between collapsed and highly extended states during extensional flow.
This ``extensional tumbling'' is driven by dynamic fluctuations in tension along chain backbones as associative bonds are driven to break and reform in the advecting network.
This is reminiscent of the tumbling observed for entangled chains in shear flow \cite{NafarSefiddashti2015IndividualFlow}, but it arises in the absence of bulk vorticity.

We simulate linear melts of $M=2000$ chains of length $N=80$ monomers using a semi-flexible Kremer-Grest bead-spring model \cite{Kremer1990DynamicsSimulation}.
Monomers interact via a purely repulsive Lennard-Jones (LJ) potential truncated at a cutoff $r_c=2^{1/6}\sigma_{LJ}$. All physical quantities are reported in reduced LJ units of length $\sigma_{LJ}$, energy $\epsilon_{LJ}$, and time $\tau_{LJ}=\sqrt{m \sigma_{LJ}^2/\epsilon_{LJ}}$. Monomers are connected into chains by FENE bonds with an average bond length $b\approx0.96\sigma_{LJ}$ \cite{Kremer1990DynamicsSimulation}. The stiffness of chain backbones is set by applying a bending potential $k_{\theta}(1-\cos\theta)$ between adjacent bonds, where $\theta$ is the angle between each pair and $k_\theta = 1.5\epsilon_{LJ}$. 
Melts were prepared and equilibrated at monomer number density $\rho=0.85\sigma_{LJ}^{-3}$ and temperature $T=1.0\epsilon_{LJ}/k_B$, following the same way in our prior study \cite{Liu2024AEquilibrium}. These conditions produce model melts with an entanglement segment length $N_e\approx28$ beads \cite{Moreira2015DirectSimulations}, an entanglement relaxation time $\tau_e\approx1980 \tau_{LJ}$ \cite{Hsu2016StaticEquilibrium, Ge2014HealingStrength,Moreira2015DirectSimulations}, and a bare Rouse time $\tau_R^0=\tau_e(N/N_e)^2\approx16163 \tau_{LJ}$. All simulations were performed using LAMMPS \cite{Thompson2022LAMMPSScales}.

Our associative melts are made by replacing 4 normal monomers on each chain with bivalent associative monomers that interact via a 3-body Tersoff potential detailed in our recent paper \cite{Liu2024AEquilibrium}.
We evenly space associative monomers along chain backbones by replacing the 10th bead in every 20-bead subsegment with an associative bead. 
Our associative potential has two key parameters: the cohesive strength of an associative dimer $U/k_BT$ and an energy barrier $E_x/k_BT$ for forming an associative trimer \cite{Liu2024AEquilibrium}.
Varying these two parameters can realize the kinetics of a variety of associative chemistries, including H-bonding networks and covalent adaptive networks.
Here we model systems with $U/k_BT=9$, $12$, $15$, or $18$, and  $E_x/k_BT = 0.5 U/k_BT$, which display dissociative-type kinetics akin to H-bonding networks.
Following our prior study, we measure the equilibrium associative bond relaxation time --- i.e., the ``sticky" relaxation time or sticker lifetime $\tau_s^{\mathrm{eq}}$ --- for each system to be $\tau_s^{\mathrm{eq}}\approx\sim 7.1\times10^3$, $9.7\times10^4$, $1.1\times10^6$, and $1.02\times10^7$ $\tau_{LJ}$. 
Our unassociative melt systems are labeled as KG, representing Kremer-Grest normal melts.

Equilibrated melts are deformed in uniaxial extensional flow at a constant Hencky strain rate $\dot{\epsilon}=\dot{L}/L$ with $L$ the length of the system along the elongating z-axis.
The transverse dimensions contract to maintain a constant volume.
Flow is simulated using generalized Kraynik-Reinelt (GKR) boundary conditions \cite{Dobson2014PeriodicFlows, Nicholson2016MolecularExtension} and by integrating the g-SLLOD equations of motion \cite{Evans1984Nonlinear-responseFlow, Daivis2006AFlows}.
The transient extensional stress $\sigma_{E}(t)=\sigma_{zz}-\frac{1}{2}(\sigma_{xx}+\sigma_{yy})$ is computed during flow, and steady-state statistics are calculated by averaging simulation data over the strain interval $\epsilon\in[4,10]$.
Strain rates are reported in terms of two dimensionless Weissenberg numbers $\textrm{Wi}_R^0=\dot{\epsilon}\tau_R^0$ and $\textrm{Wi}_R^s=\dot{\epsilon}\tau_R^s$, normalized by either the intrinsic Rouse time of the melt with no associative bonds $\tau_R^0$ or a system-specific sticky-Rouse time $\tau_R^s=\tau_s^{\mathrm{eq}}f^2$, where $f=4$ is the number of stickers per chain and $\tau_s^{\mathrm{eq}}$ is the sticker association lifetime measured for each system in equilibrium.

\begin{figure}[ht]
\includegraphics[width=8.6cm]{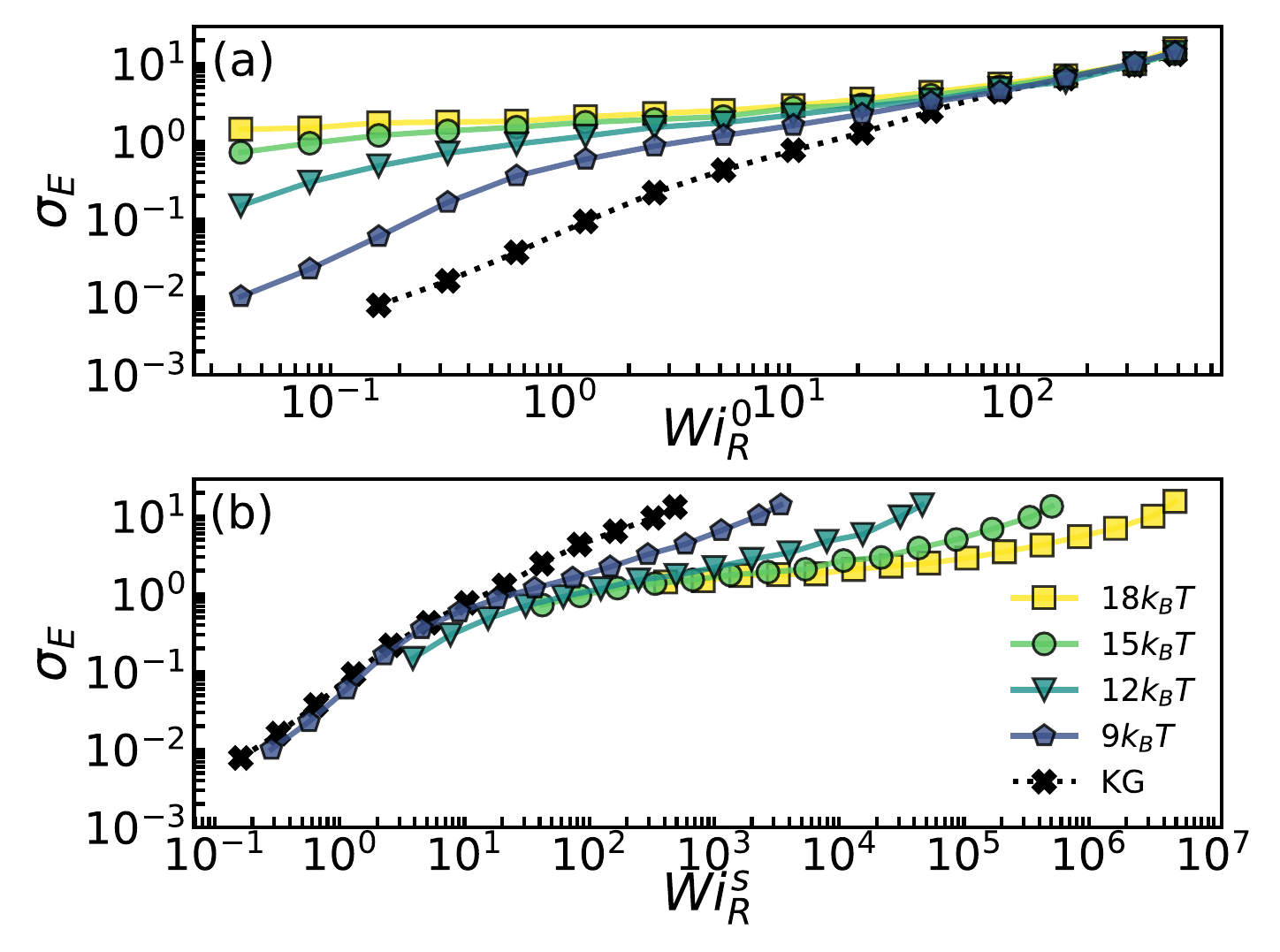}
\caption{Steady extensional stress of unassociative KG and associative melts with different cohesive strength of stickers under varying strain rates relative to (a) the original Rouse time $\tau_R^0$ for all systems, and (b) the sticky-Rouse time $\tau_R^s$ for associative systems and $\tau_R^0$ for unassociative melts.}
\label{fig:steady_stress} 
\end{figure}

Figure \ref{fig:steady_stress} shows the steady-state extensional stress $\sigma_E$ versus strain rate for dissociative-type associative and unassociative systems. 
As shown, the stresses for associative systems are significantly higher than those for the unassociative melt at low rates $\textrm{Wi}_R^0<1$, but they converge onto a common curve once $\textrm{Wi}_R^0\sim40$.
This convergence corresponds to stretched chain backbones approaching their maximum extension as shown in Fig. S1(a). In this limit, the finite extensibility of the chain backbone dominates over the chemical interactions between chains in the dissipative stress.

The rate dependence of $\sigma_E$ is controlled by the cooperative relaxation dynamics of associative bonds, which becomes clear when the stress data are replotted in Fig. \ref{fig:steady_stress}(b) with rates normalized by the sticky-Rouse time $\tau_R^s$ of each system. For the unassociative case (black X), $\tau_R^s=\tau_R^0$.
With this normalization, all systems collapse onto a common low-rate envelope that displays an initial power-law rise towards a common stress plateau once $\textrm{Wi}_R^s\sim1$.
Each system remains in this stress plateau until finite extensibility overtakes the dynamics at high rates.
The common response of systems below $\textrm{Wi}_R^s\sim1$ is consistent with the ideas of the sticky-Rouse model, but the subsequent stress plateau is unanticipated by mean-field theories. 
As we will show, this plateau is also governed by the dynamics of associations which drive chains to tumble as they fluctuate in extensional flow.

\begin{figure}[ht]
\includegraphics[width=8.6cm]{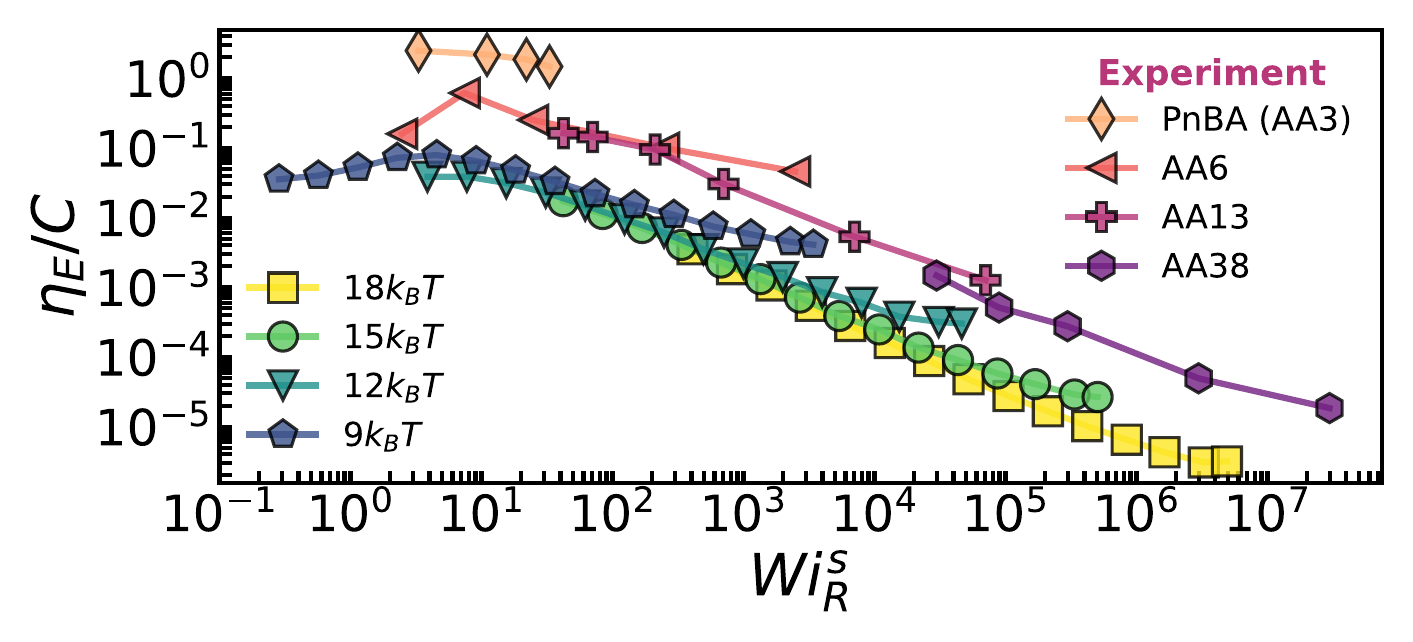}
\caption{Comparison on normalized extensional viscosity of associative melts with varying bond strengths and experiment \cite{Shabbir2015EffectMelts} on poly(n-butyl acrylate) (PnBA) and PnBA-poly(acrylic acid) H-bonding copolymers with varying percentages of acrylic acid (AA) associative groups.
Viscosities are scaled vertically by a constant $C$ for visual comparison, where $C=\tau_R^s$ and  $10^4\tau_R^s$ for simulations and experiments, respectively.}
\label{fig:steady_viscosity_btw_simu_and_exp} 
\end{figure}

MD predictions for the steady-state viscosity $\eta_E(\dot\epsilon)$ agree well with experiments.
Figure \ref{fig:steady_viscosity_btw_simu_and_exp} plots $\eta_E$ versus $\textrm{Wi}_R^s=\dot{\epsilon}\tau_R^s$ for both simulations and filament stretching experiments on acrylate-based H-bonding networks \cite{Shabbir2015EffectMelts}.
Experimental $\eta_E$ are taken from the terminal values of the published transient viscosities, and the relaxation times of their H-bonding copolymers are taken from the fitted models of Wanger et al. \cite{Wagner2022ModellingMelts}. 
Both simulations and experiments predict flow curves with near-perfect rate-thinning $\eta_E=\sigma_E/\dot{\epsilon}\sim\dot{\epsilon}^{-1}$, consistent with a rate-independent plateau in extensional stress like we observe.
Experiments and simulations also show similar signatures of finite extensibility, with systems lifting off the master curve at the highest rates.
The extent of the power-law thinning regime grows with increasing $U/k_BT$ in simulations and with increasing acrylic acid content in the experiments. 
In the context of our analysis, both strategies increase $\tau_R^{s}$ and should increase the plateau width.

\begin{figure}[ht]
\includegraphics[width=8.6cm]{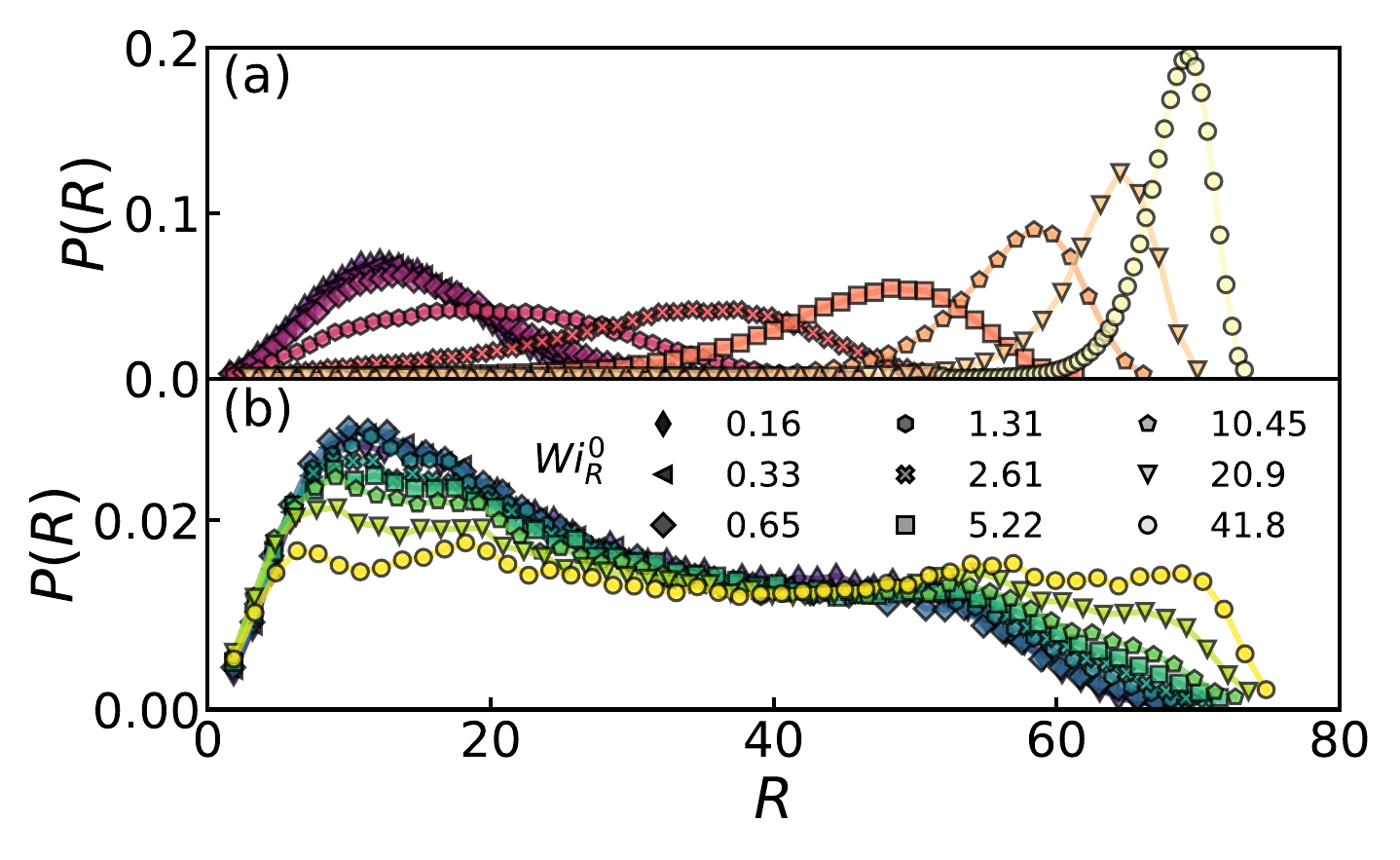}
\caption{Distribution function $P(R)$ of end-to-end distance $R$ for (a) unassociative melts and (b) associative melts with $U=18k_B T$ under uniaxial extensional flow at varying strain rates relative to the original Rouse time $\textrm{Wi}_R^0$.}
\label{fig:Ree_distribution} 
\end{figure}

To understand the molecular origin of the stress plateau, we plot the steady-state distribution of chain end-to-end vectors $P(R)$ in Fig. \ref{fig:Ree_distribution} for several $\textrm{Wi}_R^0$.
For the unassociative melt shown in Fig. \ref{fig:Ree_distribution}(a), chains elongate uniformly during extension, producing a peaked distribution that moves to larger $\langle R\rangle$ with increasing strain rate.
This is consistent with past studies \cite{OConnor2018RelatingMelts,Oconnor2019StressPolymers} and produces a monotonic rise in $\sigma_E(\dot\epsilon)$ as observed in Fig. \ref{fig:steady_stress}.
The molecular response of APN systems is very different and is shown in Fig. \ref{fig:Ree_distribution}(b) for the $U/k_BT=18$ system.
Once $\textrm{Wi}_R^s\sim1$, APN distributions flatten into a broad and almost uniform distribution that persists as the strain rate increases.
This means that the steady-state flow is populated by chains elongated to essentially all extensions in equal measure, the extreme limit of molecular-scale fluctuations.

The extensional stress $\sigma_E$ is often modeled as a function of the \textit{average} chain elongation $\langle R \rangle$, and it is common to use such models to infer molecular structure from rheological data.
This correspondence works well for normal melts, since $P(R)$ is peaked and its average captures the state of most chains.
However, for associative networks, the broad $P(R)$ results in a \textit{smaller} average chain stretch $\langle R \rangle$,
which is expected to correspond to lower macroscopic stress compared to normal melts. 
This expectation from mean-field theory is completely contradicted by the stresses shown in Fig. \ref{fig:steady_stress}(a).
Hence, the stress is not particularly meaningful for interpreting the behavior of the chains, since the average $\langle R \rangle$ is not representative of the majority of the fluid.
Additionally, such low average chain stretches counterintuitively correspond to a highly oriented average chain alignment with the flow direction as shown in Fig. S1(c).

One might ask if the wide stretch distribution in Fig. \ref{fig:Ree_distribution}(b) is due to associations trapping some chains in particularly strong associative clusters, but we observe that this is not the case.
Instead, all chains uniformly respond to flow, but their response is characterized by large fluctuations between unstretched (low $R$) and highly stretched (high $R$) states that persist in steady-state flow (See Fig. S2).
We term this behavior ``extensional tumbling'', since it resembles the end-over-end reorientation that entangled chains display in simple shear flows \cite{Schroeder2005CharacteristicFlow, NafarSefiddashti2015IndividualFlow}.
This constant cycling between extended and collapsed states produces a flat, nearly uniform distribution of $P(R)$.

While chain tumbling observed in shear is understood to be caused by the flow vorticity rotating the principal axis of elongation over time, extensional flows notably possess no net vorticity.
Thus, the ``extensional tumbling'' observed here is produced by an entirely different mechanism: stochastic fluctuations in intermolecular associative bonds along the chain.
These fluctuations give rise to tension imbalances that drive rapid relaxation in chain conformations, resulting in a dynamics very similar to chains undergoing shear.
This is an interesting instance where associative chemical interactions cause chains in extension to display chain dynamics characteristic of shear, creating a novel chemical route to tuning the nonlinear rheology of a molten polymer.

\begin{figure}[ht]
\includegraphics[width=8.6cm]{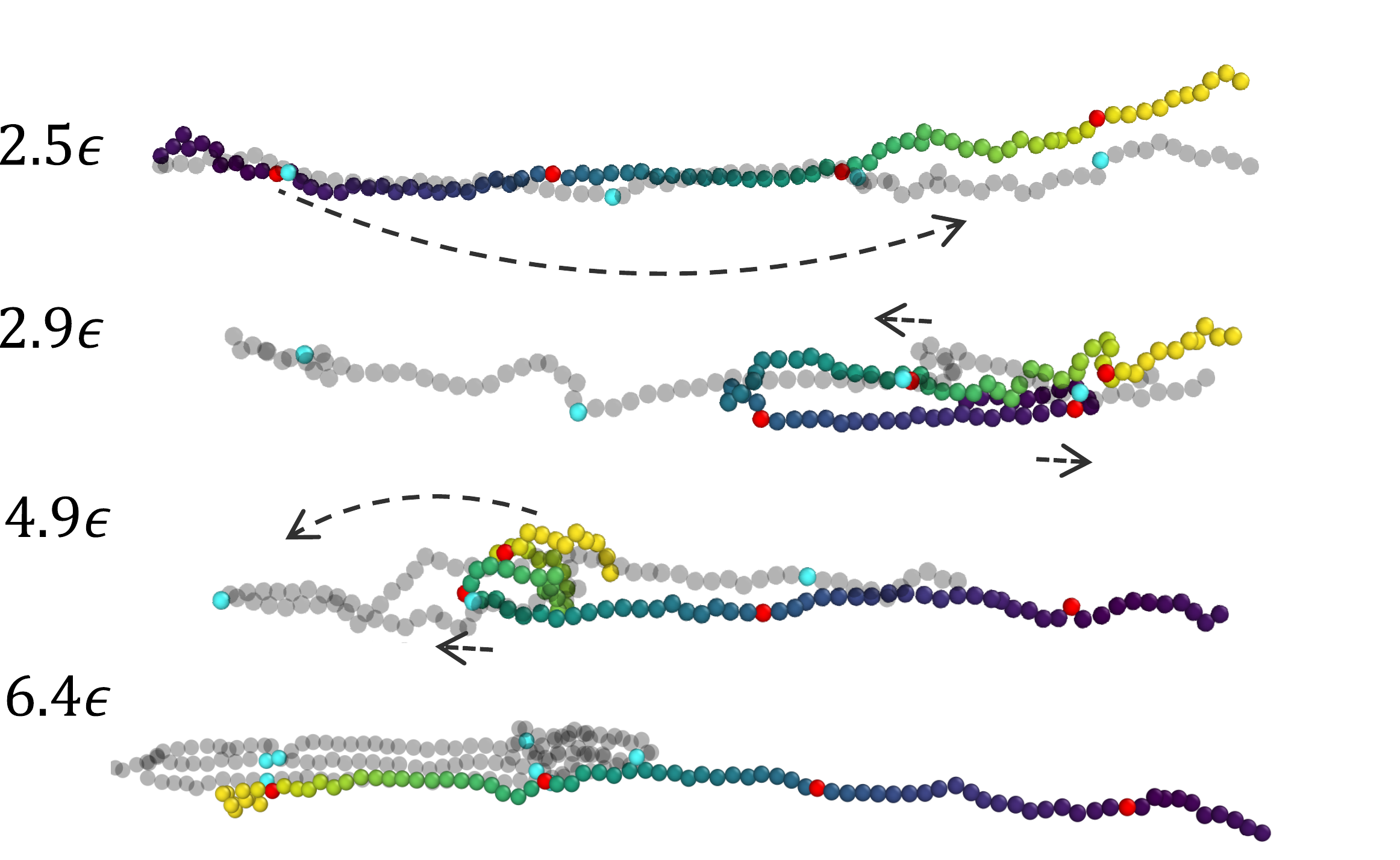}
\caption{Snapshots of a chain tumbling in uniaxial extensional flow, taken at consecutive values of strain. Beads are colored by monomer registry along the chain, with associative monomers colored red. Neighboring chains forming intermolecular associations are shown in translucent gray, with stickers in light blue. For illustrative purposes, only one or two neighboring chains critical for tumbling are included.}
\label{fig:chain_tumble} 
\end{figure}

A cluster of chains undergoing reorientation is shown in Fig. \ref{fig:chain_tumble}.
We observe that this behavior is driven by the random fluctuations of associative bonds.
As bonds break and reform, the net tension in chains fluctuates and chain segments sometimes bond to new partners moving in the opposite direction of a previous partner.
This can cause extended chains to back-fold and collapse or collapsed chains to extend as their segments are dragged around by their associative bonds.

The timescale for tumbling can be quantified by following the approach of Sefiddashti et al. for linear chains in shear \cite{NafarSefiddashti2015IndividualFlow}.
We extract a convective tumbling time $\tau_c$ by measuring the decay of the time-autocorrelation function of the unit end-to-end vector $\overrightarrow{u}$, averaged over all chains in steady state: $f(t)=\langle \overrightarrow{u}(t)\cdot \overrightarrow{u}(t+\tau)\rangle$.
We observe $f(t)$ displays a simple exponential decay $e^{-t/\tau_{c}}$ from which we extract $\tau_c$ (See Fig. S3(a)).
Notably, the pure exponential decay in $f(t)$ differs from the form observed for unassociative chains tumbling in shear, which show a decaying oscillation $f\sim \sin(\omega t)e^{-t/\tau}$ where $\omega$ and $\tau^{-1}$ are comparable in magnitude.
This is due to the fact that APN tumbling is mediated by rapid reorientation events that occur abruptly as chemical associations break and reform.

\begin{figure}[ht]
\includegraphics[width=8.6cm]{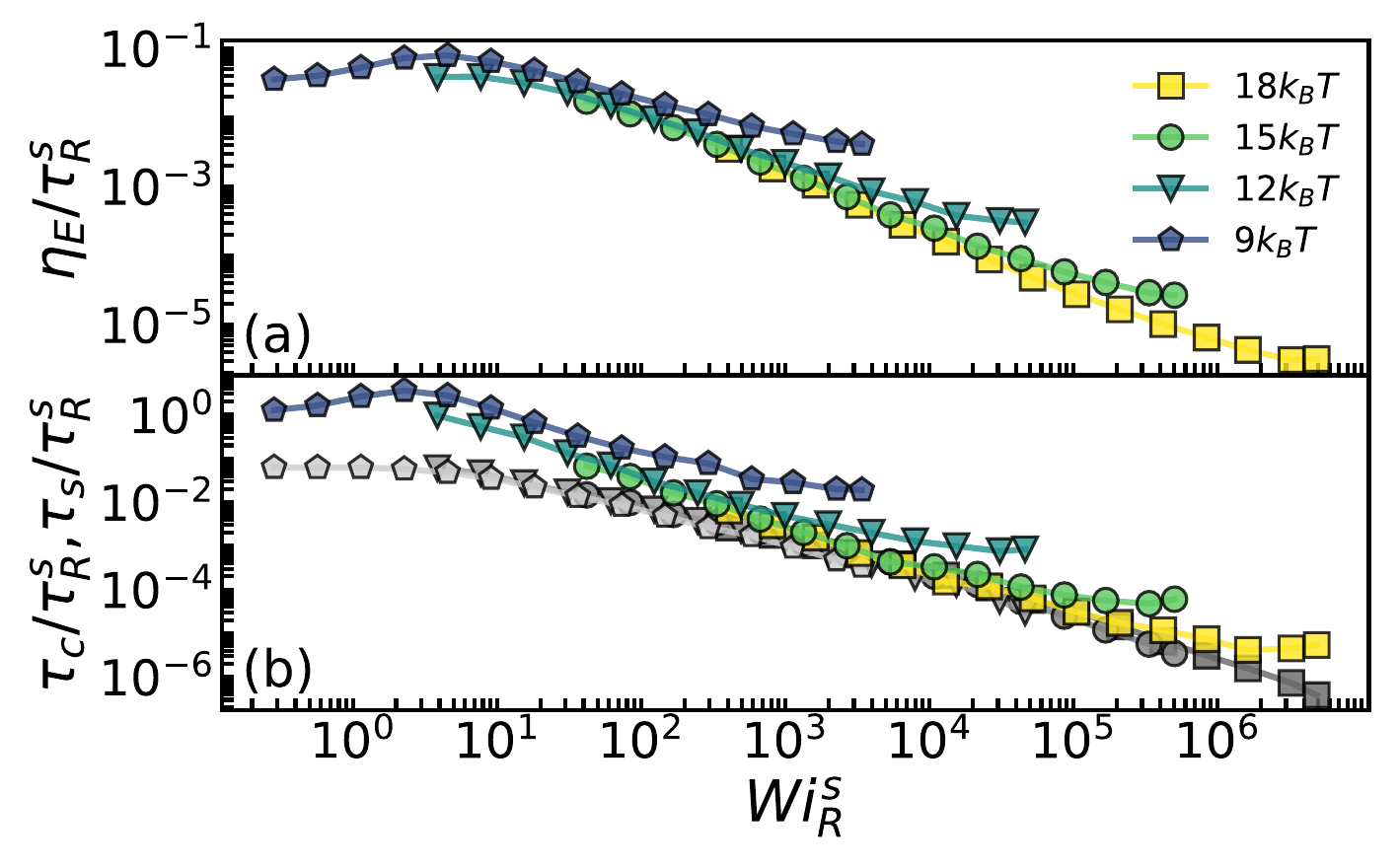}
\caption{(a) Scaled steady-state extensional viscosity $\eta_E$ and (b) scaled tumbling timescale $\tau_c$ and sticker lifetime $\tau_s$ (in gray) for associating networks with different cohesive energy of stickers. $\eta_E$, $\tau_c$, and $\tau_s$ are all normalized by sticky-Rouse timescale $\tau_R^s$.}
\label{fig:eta_tauc_taus} 
\end{figure}

Figure \ref{fig:eta_tauc_taus}(b) plots the normalized tumbling time $\tau_c/\tau_R^s$ versus Wi$_R^s$ for all four $U/k_BT$.
Normalizing by $\tau_R^s$ collapses the data onto a common curve that displays a non-monotonic shape and strongly resembles that of the viscosity $\eta_E$, replotted in Fig. \ref{fig:eta_tauc_taus}(a). 
At low Wi$_R^s<1$, the tumbling time approaches the sticky-Rouse time $\tau_c\sim\tau_R^s$ as chain reorientation is described by the sticky-Rouse model near equilibrium.
For large Wi$_R^s\gg1$, the tumbling time follows a power-law as it approaches the sticky relaxation time of individual associations $\tau_s$, plotted as gray symbols in Fig. \ref{fig:eta_tauc_taus}(b). 
In this limit, chain tumbling is limited by the speed at which individual associations reform.
At intermediate Wi$_R^s \sim1$, $\tau_c$ reaches a maximum, indicating a slowdown in the tumbling process as chains begin to elongate.

Previous studies
\cite{Vaccaro2000APolymers,Tanaka1992ViscoelasticModuli,Tanaka1992ViscoelasticTheory,Tanaka1992ViscoelasticPhenomena,Tanaka1992ViscoelasticViscoelasticity,Sing2015CelebratingNetworks} have generally agreed that the nonequilibrium kinetics of bond-dissociation should govern the rate-dependence of the APN viscosity, but the cooperative mechanisms mediating this relationship have not been established.
Figure \ref{fig:eta_tauc_taus} shows that the viscosity $\eta_E\sim\tau_c$, with both displaying a common shape, including a maximum at $Wi_R^s\sim1$.
But this maximum is notably absent in the rate-dependence of $\tau_s$.
This strongly suggests that the nonlinear dissipation is driven by chain tumbling and the non-monotonic viscosity emerges from the cooperative interaction of separate chains within the network, rather than from the kinetics of individual associative bonds.

The tumbling we observe here can be considered as a nonequilibrium generalization of chain ``brachiation.''
Cai and Spakowitz introduced the concept of ``brachiation'' by which an associative chain ``swings'' its segments through an APN, similar to how a monkey swings from branch-to-branch through a forest \cite{Cai2020BrachiationNetwork}.
Our prior simulations of the same model APNs measured chain brachiation times in equilibrium and showed that this timescale governs the terminal relaxation time of binary APNs \cite{Liu2024AEquilibrium}.
The present work now reveals that nonlinear flows drive brachiating chains to undergo dramatic fluctuations in chain stretch as chain segments swing between sections of the network advecting in opposing directions (Fig. \ref{fig:chain_tumble}).

Notably, the extensional tumbling we observe in APNs is not observed in other transient polymer networks mediated by topological entanglements \cite{OConnor2018RelatingMelts} or uncoordinated ionic aggregates \cite{Mohottalalage2022NonlinearFlow}, despite all displaying similar sticky-Rouse/Reptation linear viscoelasticity.
Instead, it seems that each form of intermolecular constraint can drive different nonequilibrium mechanisms coupling network relaxation to flow \cite{Graham2003MicroscopicRelease,Mohottalalage2022NonlinearFlow, Mohottalalage2023BeadSpringBehavior}.
Thus, while many APN chemistries display universal functional form in linear response, the precise nature, coordination, and kinetics of associative complexes appear to drive fundamentally different nonlinear dynamics.

To conclude, we have shown that a chemically generic coarse-grained model for bivalent APNs accurately predicts the experimental power-law rate-thinning of the extensional viscosity of H-bonding networks \cite{Shabbir2015EffectMelts}.
Simulations relate the strong power-law thinning of $\eta_E$ to a rate-independent plateau in both stress and average chain elongation, which develops as brachiating chains undergo a novel form of ``extensional tumbling.''
The fluctuating associations connecting a chain to the flowing network produce local tension imbalances along chain backbones, driving chains to continuously elongate and collapse.
Tumbling chains develop a broad distribution of chain stretch with a rate-insensitive average that persists until chains approach finite extensibility.
This process is consistent with the theoretical arguments made by Schaefer and McLeish for associative silks \cite{Schaefer2021PowerFlow, Schaefer2021StretchingFlow, Schaefer2022TheoreticalCrystallization}.

Despite its minimal description of associative interactions, our model shows good agreement with experiments. 
This suggests that the extensional tumbling we observe may be a generic feature of APNs with binary coordination.
We suspect that other association coordinations and kinetics may give rise to even more novel behaviors and flow sensitivities.
Our future work aims to explore how varying these parameters influences APN dynamics and rheology.
We believe that our coarse-grained models will be a valuable tool for understanding the physics of this complex chemical landscape.

\section{acknowledgments}
This work was supported by the U.S. Department of Energy, Office of Science, Basic Energy Sciences under Award \# DESC0025643.
 This work was performed, in part, at the Center for Integrated Nanotechnologies, an Office of Science User Facility operated for the U.S. Department of Energy (DOE) Office of Science. Sandia National Laboratories is a multi-mission laboratory managed and operated by National Technology \& Engineering Solutions of Sandia, LLC, a wholly owned subsidiary of Honeywell International, Inc., for the U.S. DOE’s National Nuclear Security Administration under Contract No. DE-NA-0003525. The views expressed in the Letter do not necessarily represent the views of the U.S. DOE or the U.S. Government.

\nocite{*}

\bibliography{reference}

\end{document}


\preprint{APS/123-QED}

\title{Supplemental Material:\\ Flow-driven Stretch Fluctuations Cause Anomalous Rate-Thinning In Elongating Associative Polymers}


\author{Songyue Liu, Thomas C. O'Connor}
\affiliation{%
 Department of Materials Science and Engineering, Carnegie Mellon University, Pittsburgh, PA, USA
}%

\date{\today}

\date{\today}

\maketitle

\renewcommand{\thefigure}{S\arabic{figure}}

\begin{figure}[ht]
\includegraphics[width=12cm]{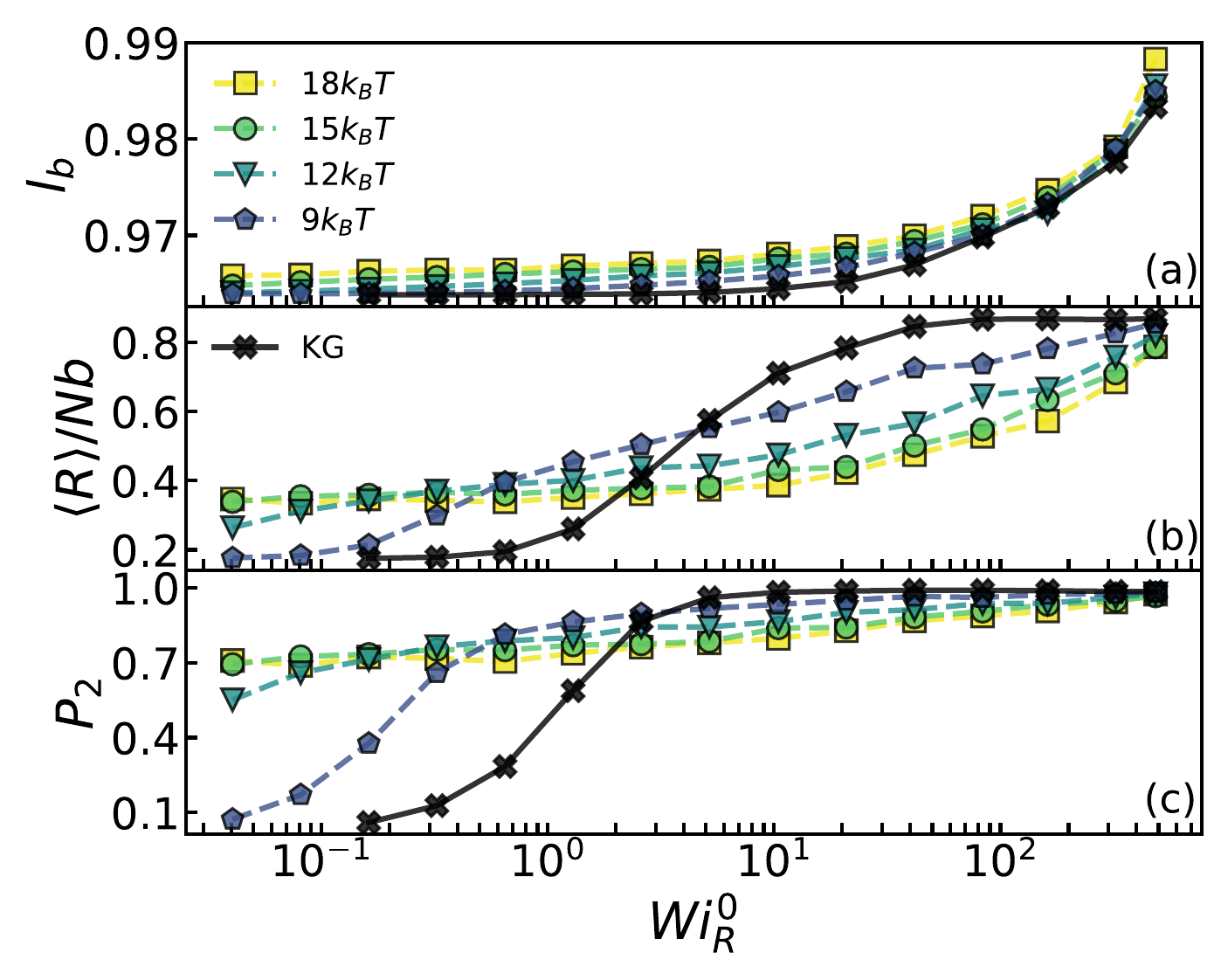}
\caption{Steady state averages for (a) bond length $l_b$, (b) end-to-end extension ratio $\langle R \rangle/Nb$, and (c) end-to-end orientational order parameter $P_2=\frac{1}{2}(3\langle\cos^2{\theta}\rangle-1)$ in both associative and unassociative (KG) melts for varying Wi$_R^0$.}
\label{fig:avg_bond_h_ee} 
\end{figure}


\begin{figure}[ht]
\includegraphics[width=12cm]{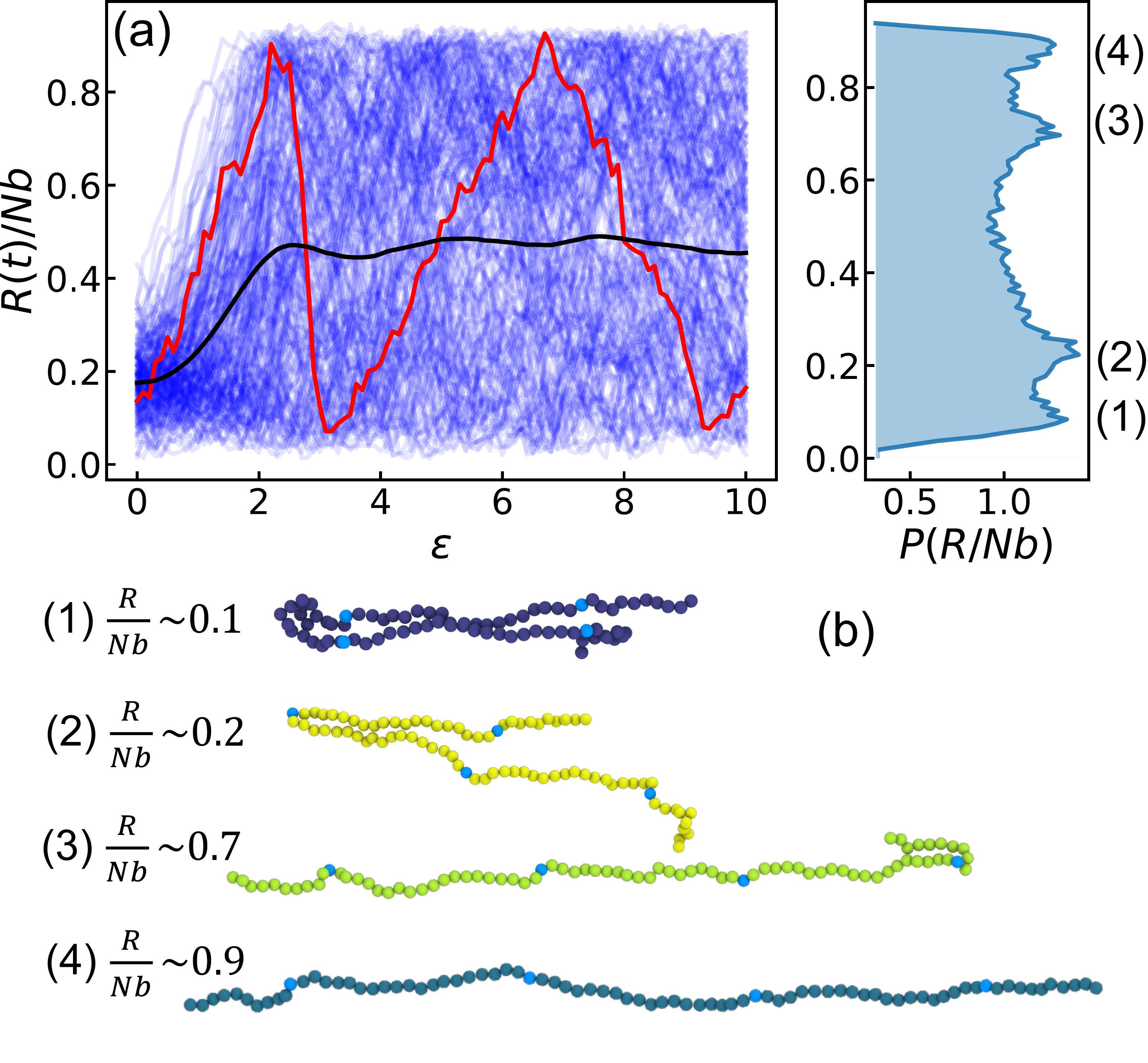}
\caption{(a) The end-to-end extension ratio $R(t)/Nb$ over strain of each chain under uniaxial extensional flow at Wi$_R^0$=41.8 for associating systems with $U/k_BT=18$. Each blue line represents the end-to-end extension of a single chain. The solid black line is the average value of $R(t)/Nb$ for all the chains. The red line highlights the $R(t)/Nb$ of one chain to show its fluctuation of extension over time. The right panel shows the average distribution $P(R/Nb)$ in steady state. (b) Illustrations of individual chain conformation associated with different peaks $P(R/Nb)$. The stickers are highlighted in light blue.}
\label{fig:h_ee_dis_and_illus} 
\end{figure}


\begin{figure}[ht]
\includegraphics[width=16.5cm]{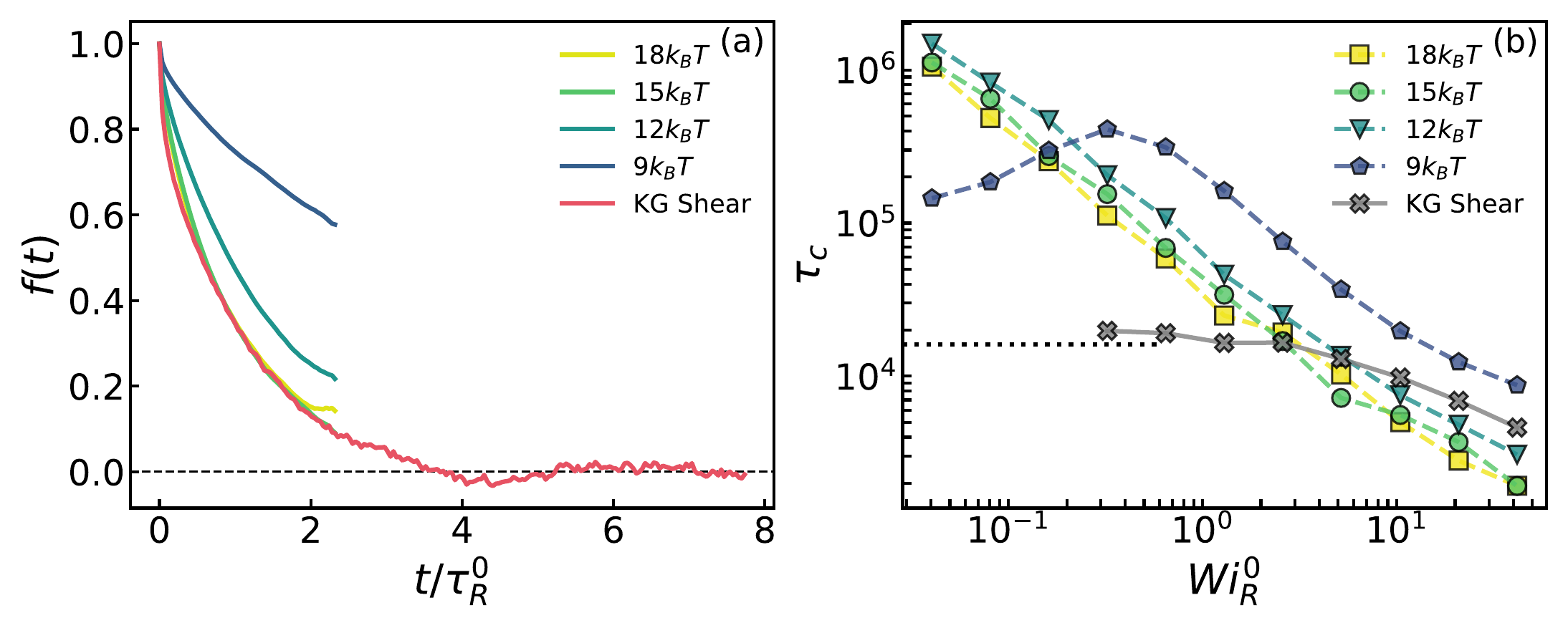}
\caption{(a) The autocorrelation function of the unit end-to-end vector $\overrightarrow{u}$ at Wi$_R^0=2.61$ for varying $U/k_BT$. (b) Steady-state tumbling timescale extracted from curves like those in (a) for varying $U/k_BT$ and Wi$_R^0$. Tumbling data for an unassociative Kremer-Grest (KG) melt during simple shear flow is shown for comparison. The dotted line indicates the equilibrium Rouse time $\tau_R^0$ for an unassociative chain.}
\label{fig:correlation_func_and_avg_tumb_time} 
\end{figure}

